\documentclass[11pt,a4paper]{article}

\usepackage{amsmath}
\usepackage{amssymb}
\usepackage{amsfonts}
\usepackage{amsthm}
\usepackage{mathtools}
\usepackage[T1]{fontenc}

\newtheorem*{theorem}{Theorem}
\theoremstyle{definition}
\newtheorem{example}{Example}

\newcommand{\dd}[1][]{\mathrm{d}{#1}}

\newcommand{\F}[1][]{\mathsf{F}_{#1}^{}}
\renewcommand{\Finv}[1][]{\mathsf{F}_{#1}^{-1}}
\newcommand{\hpsi}{\hat\psi}

\newcommand{\trunc}{{\text{trunc.}}}
\newcommand{\htwo}{{\frac \hbar 2}}
\newcommand{\dir}[1]{\delta\big( #1 \big)}
\newcommand{\ket}[1]{\left| #1 \right>}

\title{Wigner function of a qubit}
\author{Jerzy Kijowski, Piotr Waluk, Katarzyna Senger \\
Center for Theoretical Physics, Polish Academy of Sciences \\
	Al. Lotnik\'ow 32/46, 02-668 Warszawa, Poland\\
E-mail: piotr.waluk@cft.edu.pl}

\begin{document}


\maketitle

\begin{abstract}
We show that real polarization method can be effectively used to geometrically quantize physical systems with compact phase space, like the spin.
Our method enables us to construct a wave function of a qubit in both position and momentum representations and also its Wigner function. These results can be used in quantum informatics.
\end{abstract}

\section{Introduction}

Quantization of spin was often approached {\em via} a symplectic reduction of phase space of a rigid body. It seems, however, that spin of an elementary particle (e.g. an electron) has nothing to do with any classical rotation. It just describes a transversal (with respect to the four-velocity $u=(u^\mu)$) component of the energy-momentum tensor. Its possible values $\sigma=(\sigma^\mu)$ fill, therefore, a sphere $\mathbb{S}^2$ of vectors which: 1)are orthogonal to $u$ (i.e. $u^\mu \sigma_\mu = 0$) and 2) have fixed length $\sigma^\mu \sigma_\mu$, characteristic for the given particle. However, the sphere cannot be interpreted as a configuration space because different components of angular momentum do not commute. The~sphere carries a natural symplectic structure, namely the volume structure induced by spacetime metric. Due to this, it can be interpreted as a {\em phase space} of the spin.

Such a {\em compact} phase space used to be quantized in complex polarization (see e.g.~\cite{Duval}, \cite{Michel} and references therein). In this paper we show how to do it in a much more orthodox way: using real (position or momentum) polarization and following the procedure proposed in \cite{GQ-77}, \cite{geoquant} or \cite{timeoperator}. The main advantage of this approach is that it produces standard structures of quantum mechanics, like wave function (with square of its modulus describing probability), relation between position and momentum representation described in terms of the Fourier transformation and, finally, the Wigner function, which are absent in the conventional description of spin systems. We stress that these structures provide entirely new tools in the analysis of space of quantum states of such system. We very much believe in its applicability in the quantum information theory.

\section{Phase space of the spin}

Phase space ${\cal P}$ of spin $s$ is a 2-dimensional sphere $\mathbb{S}^2$ of radius $\sqrt{s}$, equipped with a natural symplectic structure, given by its volume form in one of the two possible orientations. We choose the following one:
\begin{equation}
\omega_s=-s\sin\vartheta \, \dd[\vartheta]\wedge\dd[\varphi] = \dd[\varphi]\wedge\dd[(-s\cos \vartheta)]
= \dd[\varphi]\wedge\dd[\xi]\, , \label{form}
\end{equation}
where $(\varphi,\vartheta)$ are spherical coordinates and $\xi:=s(1-\cos\vartheta)-\frac \hbar 2$. The ``geographical longitude'' $\varphi$ plays role of ``momentum'' canonically conjugate to the ``position'', described by the variable $\xi$ (cf.~formula $\omega=\dd[p]\wedge\dd[q]$ in mechanics).

${\cal P}$ can be identified with the rectangle $R_{(0,0)}:=[0,2\pi[\times[-\frac \hbar 2,2s-\frac \hbar 2[ \subset \mathbb{R}^2$ or any other rectanle $R_{(n,m)}$ obtained from $R_{(0,0)}$ using a shift by ``$2n\pi$'' in the first and by ``$2ms$'' in the second variable. For this purpose we define the following mapping:
\begin{equation}
\begin{aligned}
A:&&\mathbb{R}^2 &\to \mathbb{S}^2 \\
&&(\varphi,\xi)&\mapsto (\varphi_\trunc,\xi_\trunc) \, ,
\end{aligned}
\end{equation}
where ``truncated'' values correspond to $R_{(0,0)}$, according to formulae:
\[
\varphi_\trunc= \varphi - 2n\pi  \in [0,2\pi [ \ \ ; \ \  \xi_\trunc= \xi - 2ms \in [-\frac \hbar 2,2s-\frac \hbar 2[ \, .
\]
Map $A$ is a local diffeomorphism everywhere, with the exception of the poles of the sphere. We shall, however, ignore this discontinuity (and\dots hope for the best). Pull-back of $\omega_s$ produces the standard symplectic form on $\mathbb{R}^2$, considered as the phase space of a mechanical system with one degree of freedom. Our quantization procedure is based on the following idea: we follow a standard quantum mechanical procedure on $\mathbb{R}^2$, but restrict ourselves to quantum states which are ``the same'' in each of the sectors $R_{(n,m)}$, because such a state can be interpreted as a pull-back of the quantum state from $\mathbb{S}^2$. This ``periodicity'' condition for a quantum state will be precisely defined in the sequel.

At this point mapping $A$  can be used for quantization of both the torus $\mathbb{T}^2$ and the sphere $\mathbb{S}^2$.  The two cases are distinguished by their groups of symmetries: $T^2$ for the torus and $SO(3)$ for the sphere. We are going to quantize the latter.

We use the following convention for the Fourier transformation relating position representation $\psi$ with the momentum representation $\hpsi$ of the quantum state in quantization of $\mathbb{R}^2$:
\begin{equation*}
(\F \psi) (\varphi) = \frac 1h \int_{-\infty}^{+\infty} \dd{\xi} \, \psi(\xi) \, e^{-\frac i\hbar \varphi\xi} \qquad
(\Finv \hpsi) (\xi) =  \int_{-\infty}^{+\infty} \dd{\varphi} \, \hpsi(\varphi) \, e^{\frac i\hbar \varphi\xi} \, .
\end{equation*}

\section{A naive approach: periodicity of the wave function}

One could naively assume that ``periodicity of quantum state'' means simply ``periodicity of the wave function'' in both its position and momentum representations. Following this idea, we observe that periodicity in momentum representation implies:
\begin{equation*}
\psi(\xi)=(\Finv \hpsi) (\xi) =  \int \dd{\varphi} \, \hpsi(\varphi) \, e^{\frac i\hbar (\varphi+2\pi)\xi}=e^{\frac i\hbar 2\pi\xi}\psi(\xi) \, .
\end{equation*}
This means that $\psi(\xi)$ may assume a non-zero value only when $\xi=k\hbar, \, k\in\mathbb{Z}$. Together with periodicity of $\psi$, this implies a quantization condition for $s$:
\begin{equation}
2s=N\hbar,\quad N\in\mathbb{N} \, ,
\end{equation}
where $N$ represents the number of different values of the wave function contained within a single period $R_{(n,m)}$ i.e. the number of (complex) degrees of freedom of the system. In a similar way, we find out that $\hpsi(\varphi)$ can assume non-zero values only for $\varphi=2\pi\frac kN$, $k\in\mathbb{Z}$. Hence, these wave functions {\em do not} belong to the $L^2$ class but are distributions of the Dirac delta type:
\begin{equation}
\label{explicitWF-1}
\begin{aligned}
\psi(\xi)&=\sum_{l=-\infty}^{+\infty}\sum_{k=0}^{N-1}\psi_k \dir{\xi-(Nl+k)\hbar}, \\
\hpsi(\varphi)&=\sum_{l=-\infty}^{+\infty}\sum_{k=0}^{N-1} \frac {\hpsi_k}{\hbar\sqrt{N}}  \dir{\varphi-\frac{2\pi}N(Nl+ k)}.
\end{aligned}
\end{equation}
Using formula:
\begin{equation*}
\F \left(\sum_{ k=-\infty}^{+\infty} \delta(\xi - \alpha k) \right)=\frac 1\alpha \sum_{ k=-\infty}^{+\infty} \dir{\varphi - \frac h\alpha k},
\end{equation*}
we can derive the relation between $\psi_k$ and $\hpsi_k$:
\begin{equation}
\label{coefficientsTransformation}
\psi_k =  \sum_{m=0}^{N-1} \frac 1{\sqrt{N}} \hpsi_m e^{ 2\pi i \frac {km}N}\, ,
\qquad
\hpsi_k= \sum_{m=0}^{N-1} \frac 1{\sqrt{N}} \psi_m e^{-2\pi i \frac {km}N} \, .
\end{equation}
The space $H_N$ of these wave functions carries a natural structure of a Hilbert space, isomorphic to $\mathbb{C}^N$, where
\begin{equation*}
    \|\psi\|^2 = \sum_{k=0}^{N-1}|\psi_k|^2 = \sum_{k=0}^{N-1} |\hpsi_k|^2 = \| \hpsi \|^2 \, .
\end{equation*}

\section{Periodicity of the quantum state}

The above framework is physically unacceptable because it favors the $\varphi=0$ meridian: rotation $\varphi \longrightarrow \varphi - \varphi_0$ cannot be implemented within this picture. Indeed, such a rotation would imply:
\begin{equation}
\psi(\xi)= \int \dd{\varphi} \, \hpsi(\varphi) \, e^{\frac i\hbar \varphi\xi} \xrightarrow{\text{rotation}}  \int \dd{\varphi} \, \hpsi(\varphi-\varphi_0) \, e^{\frac i\hbar \varphi\xi}= e^{i \varphi_0\frac \xi\hbar }\psi(\xi) \, ,\label{rotation}
\end{equation}
i.e.~the function $\psi$ would no longer be periodic, due to the phase factor $e^{i \varphi_0\frac \xi\hbar }$, and could no longer be identified with a pull-back of a common wave function on $\mathbb{S}^2$. Observe, however, that the difference between the wave function contained in $R_{(n,m_1)}$ and $R_{(n,m_2)}$ is merely a constant phase factor, namely: $\exp\left(i \varphi_0 \frac{2s}\hbar(m_1 - m_2) \right)$. This means that the physical state described by these two wave functions is the same. We relax, therefore, the periodicity condition: not the wave function (an element of the Hilbert space $H_N$) but the physical state (an element of the corresponding {\em projective} Hilbert space) must be periodic.
This allows us to represent our wave function with an arbitrarily chosen $\varphi=\varphi_0$ meridian as a starting point for momentum representation. The formulae \eqref{explicitWF-1} then take the form:
\begin{equation}
\label{explicitWF-2}
\begin{aligned}
\psi(\xi)&=\sum_{l=-\infty}^{+\infty}\sum_{k=0}^{N-1}\psi_k e^{i \varphi_0\frac \xi\hbar } \delta(\xi-(Nl+k)\hbar), \\
\hpsi(\varphi)&=\sum_{l=-\infty}^{+\infty}\sum_{k=0}^{N-1}\frac{\hpsi_k}{\hbar\sqrt{N} } \delta(\varphi-(\varphi_0+\frac{2\pi}N(Nl+ k))).
\end{aligned}
\end{equation}
Note that this does not affect the relation \eqref{coefficientsTransformation}.

Within this new framework formula \eqref{rotation} becomes the definition of an operator representing a rotation in the projective Hilbert space. We stress that the ``shift'' of momentum by a constant value $\varphi_0$ is a conventional Galilei transformation. Compare this to a particle of mass $m$, whose momentum $p$ is shifted by the value $mV$ due to change of a reference frame to one of relative velocity $V$ (cf.~\cite{geoquant}).

\section{Remark concerning quantization of the torus}

A similar relaxation of the periodicity condition in momentum variable $\varphi$ could enable us to shift the position of our sphere in $\mathbb{R}^2$ in direction of the $\xi$ variable and to describe such a transformation on the quantum level in terms of a unitary operator.  The two shifts: in direction of $\varphi$ and in direction of $\xi$, generate the symmetry group of the torus $\mathbb{T}^2$, which obtains this way its (projective!) representation in the Hilbert space $H_N$. In this short note we skip this issue (partly because we do not know any physical system whose phase space carries the structure of a torus). We stress, however, that the latter shift does not belong to the symmetry group of the sphere and its quantization is incompatible with the quantization of $SO(3)$.

\section{Representation of the rotation group $SO(3)$}

Consider the group of rotations in 3-dimensional Euclidean space. It is generated by vector fields:
\begin{equation}
\vec{X}=y\frac{\partial}{\partial z}-z\frac{\partial}{\partial y}, \qquad
\vec{Y}=z\frac{\partial}{\partial x}-x\frac{\partial}{\partial z}, \qquad
\vec{Z}=x\frac{\partial}{\partial y}-y\frac{\partial}{\partial x},
\end{equation}
which, when restricted to the sphere $\{ r^2 = s = const \}$, turn out to be Hamiltonian fields with respect to the symplectic form \eqref{form}. Their generators are functions $rx$, $ry$ and $rz$ (we use the following convention: $\dd[f]=-\omega_s(\vec X_f, \cdot )$ for generation of the field $X_f$ by the observable $f$).
We choose our Poisson bracket convention so that it mirrors the commutation structure of Hamiltonian vector fields (this leads to $\{\xi,\varphi\}=-1$). The~generating functions satisfy then the following relations:
\begin{equation}
\left \{  rx, ry \right \} = -rz, \quad \left \{  ry, rz \right \} = -rx, \quad \left \{  rz, rx \right \} = -ry,
\end{equation}
in analogy with the commutators of fields $\vec X$, $\vec Y$ and $\vec Z$. We have:
\begin{align*}
rx & =s \sin\vartheta\cos\varphi \, , \\
ry & =s \sin\vartheta\sin\varphi \, , \\
rz & =s \cos\vartheta \, .
\end{align*}
To quantize these functions, we carry them over to $\mathbb{R}^2$ by means of $A$, obtaining:
\begin{align*}
f_x & :=rx \circ A = \cos (\varphi) S(\xi), \\
f_y & :=ry \circ A = \sin (\varphi) S(\xi), \\
f_z & :=rz \circ A = - \xi_\trunc +s-\frac \hbar 2,
\end{align*}
where we have introduced an auxilliary function:
\begin{equation}
S(\xi):=\sqrt{s^2-f_z{^2}} = \sqrt{\left(2s-\frac\hbar 2 -\xi_\trunc\right)\left(\xi_\trunc +\frac \hbar 2\right)} \label{es}
\end{equation}
We will now apply the Weyl quantization scheme. Because our Fourier transform between position and momentum representations has a standard form, the standard commutation relation follows: $[\hat\xi,\hat\varphi]=i\hbar$. Furthermore, all three of our generating functions factorize appropriately, so we can use the simplified formula \eqref{weylpol} from the appendix and quickly arrive at the answer:
\begin{align}
\hat f_x \psi(\xi)&= \frac 12 \left[S(\xi+\htwo)\psi(\xi+\hbar) + S(\xi-\htwo) \psi(\xi-\hbar) \right] \\
\hat f_y \psi(\xi)&= \frac 1{2i} \left[S(\xi+\htwo)\psi(\xi+\hbar) - S(\xi-\htwo) \psi(\xi-\hbar) \right] \label{fy}\\
\hat f_z \psi(\xi)&= (- \xi_\trunc +s-\frac \hbar 2)\psi(\xi) \label{fz}
\end{align}
We would like to interpret these operators as generators of ``quantum rotations''. To~support this intuition, let us take a closer look at \eqref{fz} and exponentiate the obtained operator:
\begin{equation*}
e^{\alpha\frac i\hbar \hat f_z}\psi(\xi)=e^{-\alpha \frac i\hbar (\xi_\trunc - s + \htwo)}\psi(\xi) \sim e^{-\alpha \frac i\hbar \xi} \psi(\xi).
\end{equation*}
The change $\xi_\trunc \to \xi$ in the last step makes use of our freedom to introduce a constant phase factor separately in each period. By comparing with \eqref{rotation}, the result is easily recognized as a quantum state rotated by an angle $-\alpha$.

Substituting \eqref{explicitWF-2} and \eqref{es} into \eqref{fy}, and taking $\varphi_0=0$ for simplicity, we obtain:
\begin{equation}
\begin{aligned}
\hat f_y \psi(\xi) =&\sum_{l=-\infty}^{+\infty} \sum_{k=0}^{N-1}\psi_k  \frac {i\hbar} 2
\left[ \sqrt{\Big( k+1 \Big) \Big( N-k-1 \Big)}\dir{\xi-(Nl+k+1)\hbar} \right. \\
  & \left. - \sqrt{k\Big( N - k \Big) }  \dir{\xi-(Nl+k-1)\hbar} \right] \, .
\end{aligned}
\end{equation}
As $\hat f_x$ differs from $\hat f_y$ only by the presence of the $i$ coefficient and the sign in the sum, we can easily read off its form from the result above. To compare these results with the standard physical textbook notation we substitute: $N=:2j+1$ and $k=:j-m$. By~$\ket{m}$ we will denote the wave function that possesses only one non-zero coefficient, namely $\psi_k=1$ for $k=j-m$. Rewriting our operators in this manner we obtain:
\begin{equation}
\begin{aligned}
\hat f_x \ket{m} &= \htwo \left[  \sqrt{(j+m)(j-m+1)} \ket{m-1} + \sqrt{(j+m+1)(j-m)} \ket{m+1} \right]\, , \\
\hat f_x \ket{m} &= \frac {i\hbar}2 \left[  \sqrt{(j+m)(j-m+1)} \ket{m-1} - \sqrt{(j+m+1)(j-m)} \ket{m+1} \right] \, , \\
\hat f_z \ket{m} &= m\hbar\ket{m} \, .
\end{aligned}
\end{equation}
The standard representation of the orthogonal group corresponding to spin $j$ has been exactly reproduced! This method can also be applied to any \mbox{$N=(2j+1)$-level} quantum system.

Above representation of generators of the group $SO(3)$ can be integrated to a representation of the entire group. For integer spin ($j=1,2, \dots$) we obtain this way a unitary representation of the group in the Hilbert space $H_N$. For half-integer spin ($j=\frac12 , \frac32, \dots$) we obtain a projective representation which can be lifted to a unitary representation of the covering group $SU(2)$.

It may not be immediately obvious, but our positioning of the sphere $\mathbb{S}^2$ in the $\xi$ variable was crucial for the Weyl quantization. As already mentioned in Section 5, the~shift in direction of the variable $\xi$ can  be defined on both the classical and quantum level. However, the simple quantization {\em via} the Weyl procedure works only when the support of $\psi(\xi)$ is arranged symmetrically in the interval between the poles of the sphere.

\section{Wigner function}

Having at our disposal not only the standard ``position representation'' $\psi$ of the quantum state, but also its momentum representation $\hpsi$, we can construct the Wigner function for this system. Let us recall the standard formula ($M$~and $\hat M$ are normalisation factors):
\begin{align*}
W(p,q):&=\hat M\int \overline{\hpsi}(p+\eta) \hpsi(p-\eta)e^{-\frac i\hbar 2q\eta} \dd[\eta] \\
&= M \int \overline{\psi}(q+\eta) \psi(q-\eta)e^{\frac i\hbar 2p\eta} \dd[\eta].
\end{align*}
Because our wave functions $\psi$ and $\hpsi$ are distributions, the above has to be read in the distributional sense. This means, that for every test function $\Phi(\varphi,\xi)$ we have:
\begin{equation}
\left< W(\varphi,\xi), \Phi(\varphi,\xi) \right> := \hat M \int \overline{\hpsi}(\varphi+\eta) \hpsi(\varphi-\eta)e^{-\frac i\hbar 2\xi\eta}  \Phi(\varphi,\xi) \dd[\eta] \dd[\varphi] \dd[\xi].
\end{equation}
This implies the following explicit formula for the distribution $W$:
\begin{equation}
\label{wigner}
\begin{aligned}
W(\varphi, \xi)&= \sum_{x,y=-\infty}^{+\infty} \sum_{k=0}^{N-1}  \frac {1}{2N}\overline{\hat \psi}_k \hat\psi_{x-k} e^{- i \pi \frac yN(2k-x)} \delta(\varphi-(\varphi_0 +\frac \pi N x)) \delta(\xi - \frac \hbar 2 y) \\
&= \sum_{x,y=-\infty}^{+\infty} \sum_{k=0}^{N-1} \frac 1{2N}\overline  \psi_k \psi_{y-k} e^{ i \pi \frac xN (2k-y)} \delta(\varphi-(\varphi_0 +\frac \pi N x)) \delta(\xi - \frac \hbar 2 y)
\end{aligned}
\end{equation}
An indexing convention $\psi_k=\psi_{k+N}$ is used here for the wave function coefficients.

We can see that our Wigner function has a form of a two-dimensional array of Dirac deltas, periodic in both $\varphi$ and $\xi$ with periods $2\pi$ and $2s$. Observe that, in contrast to the wave function, the Wigner function is strictly periodic on $\mathbb{R}^2$, even though we have kept an arbitrary $\varphi_0$ in the formula.

Recall the marginal property of the standard Wigner function:
\begin{equation*}
\int_{-\infty}^{+\infty}W(x,p)\dd[p] = \big|\psi(x) \big|^2 \ \ ; \ \  \int_{-\infty}^{+\infty}W(x,p)\dd[x] = \big|\hpsi(p) \big|^2
\, .
\end{equation*}
The integrals would turn out to be divergent, of course, if we tried to apply the above formulae to our situation.
However, $W$ can be considered as a distribution on the phase space $\mathbb{S}^2$.
This point of view suggests the following normalisation (already applied in~\eqref{wigner}): the integral of $W(\varphi,\xi)$ over $\mathbb{S}^2$ must be equal to $1$. The marginal properties are now satisfied, in an appropriately adjusted form:
\begin{equation}
\begin{aligned}
\int_0^{2\pi}W(\varphi,\xi)\dd[\varphi] &=
\sum_{k=0}^{N-1}\big|\psi_k \big|^2 \delta(\xi-k\hbar) \\
\int_{-\htwo}^{2s-\htwo}W(\varphi,\xi)\dd[\xi] &=
\sum_{k=0}^{N-1}\big|\hpsi_k\big|^2 \delta(\varphi-(\varphi_0+\frac{2\pi}Nk)_\trunc)
\, .
\end{aligned}
\end{equation}

\section{Rotation-averaged Wigner function}

The construction presented above depends upon a choice of poles (the $\xi$ axis) and a $ \varphi_0$-meridian on the sphere $\mathbb{S}^2$. This arbitrary choice is, of course, unphysical. Having already at our disposal the representation of the group $SO(3)$ constructed in Section 5, we can average the result over all possible choices, i.e.~over the entire group. Let, therefore, $\gamma \in SO(3)$ be a rotation that moves the initial distinguished axis and meridian into the new position and denote by $\mathcal{G}_\gamma$ the corresponding diffeomorphism of the sphere. Consider now the quantized version $\hat\gamma$ of such a rotation and define
\begin{equation}
\label{auxillaryWig}
F(\varphi,\xi,\gamma)= \mathcal{G}_\gamma^*W[\hat \gamma^{-1}\psi](\varphi,\xi) \, .
\end{equation}
Observe that the right hand side is well defined also in case of a half-integer spin, when the operator $\hat\gamma$ is defined up to a sign. This is due to the fact that $W$ is a hermitian product of $\psi$ and $\bar{\psi}$, which kills the sign ambiguity.

Function $F$ should represent the same quantum state $\psi$, because rotation $\mathcal{G}_\gamma$, applied on the classical level, should annihilate the rotation $\hat\gamma^{-1}$, which was applied on the quantum level to the state $\psi$. Following this idea we define the averaged Wigner function as:
\begin{equation}
\tilde W (\varphi,\xi) := \int_{SO(3)} F(\varphi,\xi,\gamma) \dd[\gamma]
\label{averaging}
\end{equation}
This integration smears the array of Dirac deltas into a smooth function on $\mathbb{S}^2$, which is a coordinate-independent representation of our quantum state - as can be easily seen from \eqref{auxillaryWig}, the averaged Wigner function possesses the covariance property:
\begin{equation*}
\mathcal{G}_\gamma^*\tilde{W}[\psi]=\tilde{W}[\hat\gamma\psi] \, .
\end{equation*}
The fact that we would obtain the same $\tilde W$ if we initially represented our quantum state in a different coordinate system follows as a direct corollary.

Similarly as in conventional quantum mechanics, Wigner function can also be assigned to an arbitrary mixed state: the mixture of quantum states is represented by the corresponding convex combination of Wigner functions.

\begin{theorem}
Averaged Wigner function maps linearly the cone of mixed states (positive, self-adjoint operators in $H_{2j+1}$ with trace equal 1) into the space of at most $2^{2j}$-pole spherical functions on $\mathbb{S}^2$. The mapping is invertible, i.e.~the averaged Wigner function contains the entire information about the corresponding quantum state.
\end{theorem}

\begin{example} For $j=\frac 12$ mixed states are described by positive, hermitian $(2\times 2)$ matrices with trace equal one. Hence, they fill a (positive) cone in a 3-dimensional real space.  The corresponding averaged Wigner functions contain only monopole and dipole components,
but the monopole enters with fixed coefficient, due to normalization condition. What remains is the 3-dimensional cone of dipole functions, restricted by the positivity condition.
\end{example}
\begin{example}
For $j=1$ the space of hermitian $(3\times 3)$ matrices with trace one is \mbox{8-dimensional}. The corresponding space of Wigner functions is also $8=3+5$ dimensional, where 3 stands for dipoles and 5 for quadrupole functions. The monopole component always enters with the same, normalized coefficient, whereas the remaining components are still restricted by the positivity condition.
\end{example}
\begin{example}
In case of $j=\frac 32$, 7 octupole functions enter the game, which rises the number of free parameters to $3+5+7=15$. Again, this corresponds to the number of independent hermitian $(4\times 4)$ matrices with trace one.
\end{example}

The fact that higher multipoles ``do not fit'' into a small sphere expresses the Heisenberg uncertainty principle.

\section*{Acknowledgments}
This work was supported in part by
Narodowe Centrum Nauki (Poland) under Grant No. DEC-2011/03/B/ST1/02625. One of the authors (P.W.) was also supported by a special internal Grant for young researchers, provided by Center for Theoretical Physics, PAS.

\newpage

\appendix
\section{Weyl quantization}
We use the following convention for the Weyl quantization:
\begin{equation}
\label{weyl}
\hat f \psi(q) := \frac 1{h^2}\int \dd[\alpha] \dd[\beta] \tilde f (\alpha,\beta)\left[ e^{\frac i\hbar (\alpha \hat q + \beta \hat p)} \psi(q) \right],
\end{equation}
where $\tilde f(\alpha,\beta)$ is the Fourier transform of the function $f(q,p)$ defined on the phase space:
\begin{equation*}
\tilde f(\alpha, \beta):=\int \dd[q] \dd[p] f(q,p) e^{-\frac i\hbar (\alpha  q + \beta  p)} \, ,
\end{equation*}
whereas $\hat q$ and $\hat p$ are quantum position and momentum operators.

Suppose now that $f$ factorizes: $f(q,p) = f_q (q)f_p (p)$. Consequently, also its Fourier transform factorizes: $ \tilde f (\alpha, \beta) = \tilde f_q (\alpha) \tilde f_p (\beta)$. Using the Campbell-Baker-Hausdorff formula we can rewrite \eqref{weyl} in the following way:
\begin{equation}
\label{weylpol}
\hat f \psi(q)= \int\dd[\beta]\frac 1h \tilde f_p(\beta)f_q(q+\frac12\beta)\psi(q+\beta) \, .
\end{equation}

\end{document}